\documentclass[preprint,12pt]{elsarticle}
\usepackage[english]{babel}
\usepackage[letterpaper,top=2cm,bottom=2cm,left=3cm,right=3cm,marginparwidth=1.75cm]{geometry}
\usepackage{amsmath}
\usepackage{graphicx}
\usepackage{float}
\usepackage[colorlinks=true, allcolors=blue]{hyperref}
\usepackage{subcaption}
\journal{Chaos, Solitons and Fractals}
\usepackage{booktabs} 
\usepackage{siunitx}  
\usepackage[table]{xcolor} 

\begin{document}

\begin{frontmatter}

\title{Experimental observation of chaotic and multistable dynamics in a Duffing--Holmes--type analog circuit: antiperiodicity and attractor-coexistence signatures}

\author{
  Patricia R. Gargiulo$^{1}$, Carac\'e Guti\'errez$^{1}$, Juan P. Tarigo$^{1}$,
  Cecilia Stari$^{1}$ and  \\ Arturo C. Mart\'i$^{1}$ \\
  $^{1}$Instituto de F\'isica, UdelaR, 11400, Montevideo, Uruguay
}
\date{}

\begin{abstract}
An experimental study of a periodically forced Duffing--Holmes-type oscillator with
a double-well potential, emulated by a piecewise-linear analog electronic circuit, is presented. By systematically varying the forcing amplitude and frequency, the full dynamical landscape of the system is characterized through bifurcation diagrams, Poincar\'e maps, and largest Lyapunov exponent calculations. The observed
phenomenology includes period-doubling routes to chaos, periodic windows with
multistability, intermittency, and antiperiodic orbits in which the trajectory
recovers the global symmetry of the double-well potential. Multistability-induced
discontinuities in the bifurcation diagrams are identified and interpreted as
attractor-coexistence signatures arising from the sensitivity of the long-time
dynamics to initial conditions, rather than as noise artifacts. These results are
synthesized into a high-resolution two-dimensional map of the parameter space. The close agreement among all experimental diagnostics validates the fidelity of this analog implementation and demonstrates that continuous-time
hardware provides a high-throughput platform for mapping complex nonlinear
landscapes and resolving fine-scale multistable structures.
\end{abstract}

\begin{keyword}
Duffing--Holmes oscillator \sep analog circuit \sep bifurcation \sep chaos \sep antiperiodicity \sep parameter-space diagram
\end{keyword}

\end{frontmatter}
\section{Introduction}

Nonlinear electronic circuits have long served as powerful experimental platforms for the investigation of complex dynamical phenomena, including bifurcations, multistability, and deterministic chaos \cite{rulkov1996images,ogorzalek1997chaos,matsumoto2003chaotic,pisarchik2005homoclinic,samardzic2017analysis}. Their appeal stems from a unique combination of features: ease of implementation, precise parameter control, real-time access to phase-space trajectories, and the natural incorporation of physical effects such as noise, component tolerances, and parasitic couplings that are often neglected in numerical studies. As a result, analog electronic systems provide an ideal bridge between mathematical models and physical reality, allowing theoretical predictions to be tested under realistic experimental conditions.

A paradigmatic example is Chua's circuit \cite{matsumoto2003chaotic}, which demonstrated that a remarkably simple electronic architecture can sustain deterministic chaotic oscillations. Since then, a wide variety of nonlinear oscillators have been successfully implemented in hardware. Beyond the Chua family, the Rössler system has received considerable attention through both numerical and experimental realizations. For example, Pisarchik and Jaimes-Reátegui reported a piecewise-linear Rössler-like circuit exhibiting homoclinic orbits, period-adding bifurcations, symmetry breaking, phase bistability, and homoclinic chaos, with excellent agreement between experiments and theoretical predictions \cite{pisarchik2005homoclinic}. More recently, analog electronic implementations have also been employed to investigate multiscroll chaotic attractors. In particular, Echenausía-Monroy \textit{et al.} developed an electronic platform capable of controlling the number of generated scrolls through a single parameter while simultaneously exhibiting bistable dynamics and complex attractor organization \cite{ECHENAUSIAMONROY2020100929}.

Since the pioneering works of the 1970s~\cite{ueda1980} and subsequent 
reviews~\cite{holmes1979nonlinear, lakshmanan1996chaos}, the Duffing 
oscillator has become a cornerstone of nonlinear dynamics and chaos 
theory. Described by a second-order differential equation with a cubic 
restoring term, it exhibits a remarkably rich phenomenology, periodic 
oscillations, multiple bifurcation routes, intermittency, and 
deterministic chaos~\cite{venkatesan2000intermittency}, with 
applications ranging from mechanical engineering and fault diagnosis in 
rotating machinery to secure communications via chaotic 
synchronization~\cite{natiq2023enhancing} and the modelling of biological 
systems~\cite{srebro1995duffing, lakshmanan2007bifurcations}. The 
Duffing--Holmes variant~\cite{moon1979magnetoelastic}, characterized by a 
double-well potential, has been extensively studied through both numerical 
and analytical approaches~\cite{holmes1979nonlinear, aledealat2019dynamics},
 including specialized mapping algorithms \cite{ontanon2025algorithm},
and interest in the system remains active, with recent developments 
including fractional-order nonlinearities~\cite{syta2014chaotic} and 
high-efficiency bistable device design~\cite{caetano2025nonlinear}.

Despite this richness of theoretical and numerical work, the literature 
still lacks systematic experimental investigations that go beyond the 
identification of individual chaotic or periodic regimes and provide 
instead a high-resolution, data-driven classification of the full 
dynamical landscape. Analog electronic implementations are particularly 
well suited to address this gap: they bridge mathematical models and 
physical reality by naturally embedding  physical imperfections such as noise, parameter tolerances, and parasitic couplings that numerical simulations typically ignore, while allowing direct reconstruction of phase-space trajectories and real-time observation of phenomena such as intermittency \cite{gutierrez2020observation, gutierrez2025non}. 
This latter behavior, which arises near boundary crises where a chaotic 
attractor collides with an unstable periodic orbit~\cite{grebogi1987critical}, 
is especially difficult to capture reliably in purely computational 
studies due to its extreme sensitivity to parameter variations.

In this work, a comprehensive experimental study of piecewise-linear Duffing--Holmes--type analog circuit is presented. 
High-resolution bifurcation diagrams are constructed to reveal the period-doubling 
route to chaos and the emergence of periodic windows. Dynamical regimes are classified directly from simultaneous voltage and inductor current 
measurements via Poincar\'e maps. These results are synthesized into a fully measured 
two-parameter diagram in the driving amplitude--frequency plane, 
providing a global and quantitative portrait of the system's dynamical 
richness. The paper is organized as follows. Section \ref{sec:model} describes the 
circuit implementation and its mathematical model. Section~\ref{sec:results} presents 
the experimental results, and Section~\ref{sec:conclusions} summarizes the conclusions and 
outlines directions for future work.

\section{The Duffing--Holmes oscillator: model and experimental implementation}
\label{sec:model}

This section describes the theoretical model and its analog electronic 
implementation. A nonlinear oscillator subject to periodic external forcing in the presence of dissipation is studied. The restoring force of the system incorporates both linear and nonlinear contributions, giving rise to a double-well potential structure.
The interplay between the nonlinear 
restoring force, damping, and periodic driving produces the rich 
dynamical behavior that is the focus of this work.

\subsection{The Duffing--Holmes oscillator and the double-well potential}

The Duffing--Holmes oscillator describes the motion of a dissipative 
particle in a double-well potential under periodic external forcing. 
Its equation of motion is

\begin{equation}
\ddot{x} + \delta\dot{x}  -\alpha x + \beta x^3 = \gamma \sin(\omega t),
\label{ec.duff}
\end{equation}
where $x$ represents the particle position, $\delta$ is the damping 
coefficient, and $\gamma$ and $\omega$ are the amplitude and frequency 
of the external forcing, respectively. The restoring force
\begin{equation}
    f(x) = \alpha x - \beta x^3
    \label{eq:restoring}
\end{equation}
combines a linear term with negative stiffness $-\alpha$ ($\alpha > 0$), which destabilizes the origin, and a nonlinear term with positive
stiffness $\beta$ ($\beta > 0$), which provides confinement at large
displacements. This force derives from the double-well potential
\begin{equation}
V(x) = -\frac{\alpha}{2}x^2 + \frac{\beta}{4}x^4,
\label{ec.pot_duff}
\end{equation}
which has an unstable maximum at the origin and two symmetric stable 
minima at $x = \pm\sqrt{\alpha/\beta}$. 
The unforced, undamped system ($\gamma = \delta = 0$) possesses three fixed points: an unstable saddle at the origin and two centers within each potential well. When damping is introduced ($\delta > 0$), these centers transition into stable foci, which then act as attractors of the unforced dissipative dynamics.

\subsection{Experimental setup}
\label{sec.montaje}

\begin{figure}[H]
  \centering
  \includegraphics[width=.5\linewidth]{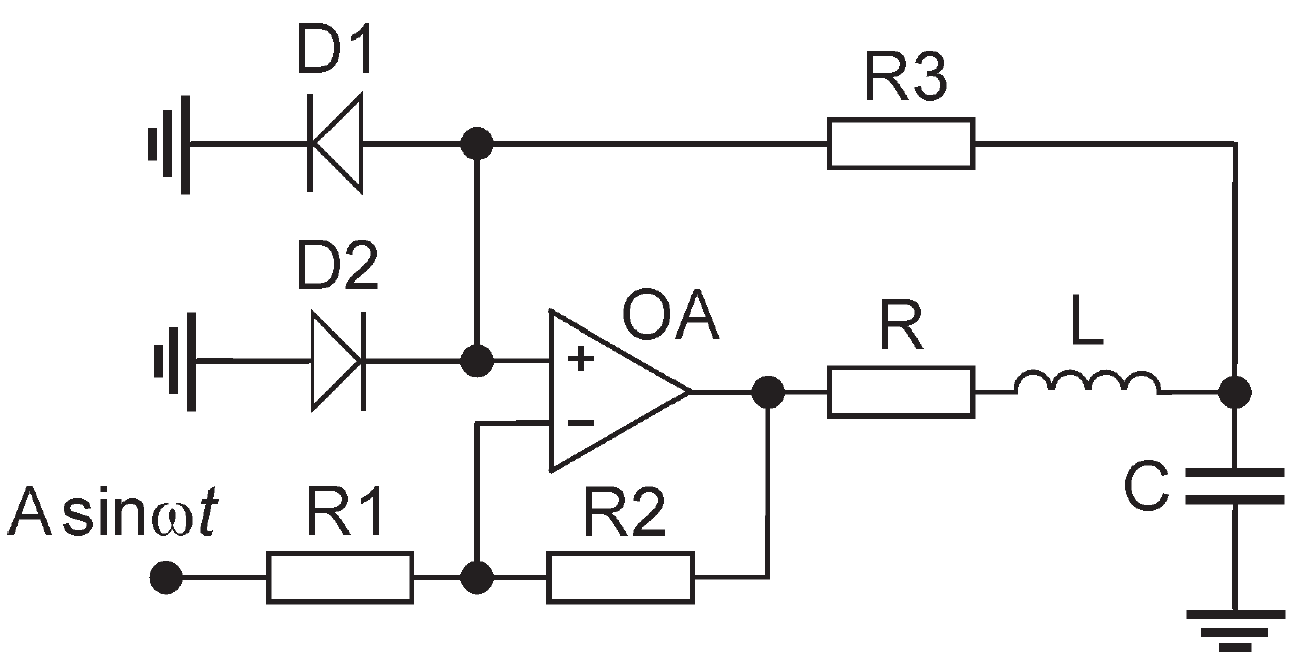}
  \caption{Schematic of the implemented circuit. The nonlinear 
  restoring force is provided by a diode array in antiparallel configuration~\cite{tamaseviciute2008analogue}.}
  \label{fig:circuito}
\end{figure}

The experimental setup consists of an analog electronic circuit that 
implements the Duffing--Holmes oscillator, whose schematic is depicted 
in Fig.~\ref{fig:circuito}. The circuit is designed so that the 
dynamical variables of the system are mapped onto measurable electrical 
quantities, enabling a direct correspondence between the physical 
oscillator and its electronic analog. The core of the circuit is a 
series $RLC$ network, in which the capacitor voltage $V_C$ and the 
inductor current $I_L$ serve as the state variables encoding the 
instantaneous phase-space coordinates of the oscillator. The passive 
elements are a $470$~nF capacitor, a $22.3$~mH inductor, and a 
variable resistor set to $R = 20~\Omega$, the latter controlling the 
effective dissipation of the system. The nonlinear restoring force 
characteristic of the double-well Duffing--Holmes potential is 
synthesized by an active subcircuit comprising an operational amplifier 
and an antiparallel diode array, which together produce a 
piecewise-linear voltage--current relationship that closely approximates 
the cubic nonlinearity of the target potential. The gain and operating 
point of this active stage are set by three fixed resistors $R_1$, 
$R_2$, and $R_3$, each of $10~\text{k}\Omega$. The resulting circuit 
topology and component values are consistent with the analog 
implementation proposed in~\cite{tamaseviciute2008analogue}.

System control and monitoring are performed using a National Instruments 
(NI) USB~6216 16-bit data acquisition card, which serves the dual 
purpose of generating the periodic forcing signal $A\sin(\omega t)$ and 
recording the oscillator's response. The interface provides an amplitude 
resolution of $3.5~\mu$V, ensuring precise control of the forcing 
parameters. The signals $V_C(t)$ and $I_L(t)$ are recorded at a 
sampling frequency of $100$~kHz, yielding time series of 
$2\times10^4$ points per channel. To ensure that observations correspond 
to the stationary regime and to avoid transient effects arising from the 
initial charging of the components, the first $10^4$ points of each 
series are systematically discarded prior to analysis.

\subsection{Circuit modeling}

The dynamics of the implemented circuit shown in Fig.~\ref{fig:circuito} can 
be analyzed using Kirchhoff's laws. Assuming $R_3 \gg \rho = \sqrt{L/C}$, 
the system is described by the coupled differential equations

\begin{equation}
\begin{aligned}
C \frac{dV_C}{dt} &= I_L, \\
L \frac{dI_L}{dt} &= F_E(V_C) - I_L R + A \sin(\omega t).
\end{aligned}
\label{eq:duffing}
\end{equation}
The nonlinear function $F_E(V_C)$, generated by the diode array, is 
modeled by a three-segment piecewise-linear approximation:
\begin{equation}
    F_{E}(V_{C}) =
    \begin{cases}
      -(V_{C} + k V^{*}), & V_{C} < -V^{*}, \\
      (k - 1)V_{C},       & -V^{*} \le V_{C} \le V^{*}, \\
      -(V_{C} - k V^{*}), & V_{C} > V^{*},
    \end{cases}
    \label{ec.lineal_a_tramos}
\end{equation}
where $k = R_2/R_1 + 1$ is the gain of the amplification stage and 
$V^* \approx 0.50$~V is the diode threshold voltage. Setting $R_2 = R_1$ 
yields $k = 2$, which is adopted throughout this work. The model 
assumes ideal diodes satisfying $R_{d0} \gg R_3 \gg R_{d1}$, where 
$R_{d0}$ and $R_{d1}$ are the blocking and conducting resistances, 
respectively.

Introducing the dimensionless variables and parameters as proposed 
in~\cite{tamaseviciute2008analogue},
\[
x = \frac{V_C}{2V^*}, \quad y = \frac{\rho\, I_L}{2V^*}, \quad
\frac{t}{\sqrt{LC}} \to t, \quad \omega\sqrt{LC} \to \omega,
\]
\[
a = \frac{A}{2V^*}, \quad b = \frac{R}{\rho}, \quad 
\rho = \sqrt{\frac{L}{C}},
\]
the circuit equations reduce to the dimensionless system
\begin{equation}
\begin{aligned}
\dot{x} &= y,  \\
\dot{y} &= F_E(x) - by + a \sin \omega t,
\end{aligned}
\label{ec.adimensionada}
\end{equation}
which is analogous to the Duffing--Holmes equation~\eqref{ec.duff}; 
here $F_E(x)$, given by Eq.~\eqref{ec.lineal_a_tramos}, is a 
piecewise-linear approximation to the cubic derivative of the 
double-well potential $V(x) = -\tfrac{1}{2}x^2 + \tfrac{1}{4}x^4$. Depending on the 
driving amplitude $a$ and frequency $\omega$, the trajectory may 
remain confined within one potential well, exhibiting periodic 
oscillations, or acquire sufficient energy to cross the potential 
barrier and jump between the two wells. This competition between 
dissipation and external excitation underlies the period-doubling 
cascade and the formation of strange attractors reported in the 
following section.

\section{Results and discussion}
\label{sec:results}

The experimental results are presented through four complementary analyses.
The main dynamical regimes of the system are first illustrated by
representative time series, phase portraits, and Poincar\'e maps. The global
evolution of the dynamics is then examined through bifurcation diagrams and
largest Lyapunov exponent calculations as each control parameter is varied
independently. A unified portrait of the dynamical landscape is provided by a
high-resolution two-dimensional map of the driving amplitude--frequency
parameter space. Finally, the antiperiodic orbits that emerge within periodic
windows, a distinctive feature of systems with symmetric double-well
potentials, are analyzed.

\subsection{Time series and Poincar\'e maps}
\begin{figure}[hbtp]
\captionsetup[subfigure]{font=normalsize}
\begin{subfigure}{1\textwidth}
    \centering
    \caption{}
    \includegraphics[width=0.45\linewidth]{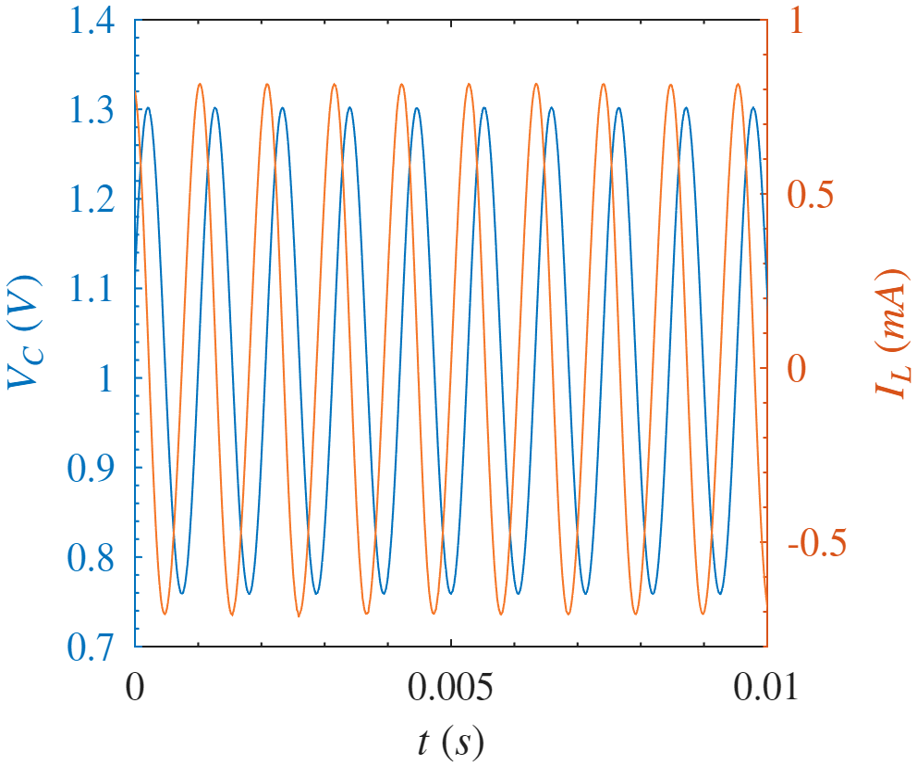}
    \includegraphics[width=0.45\linewidth]{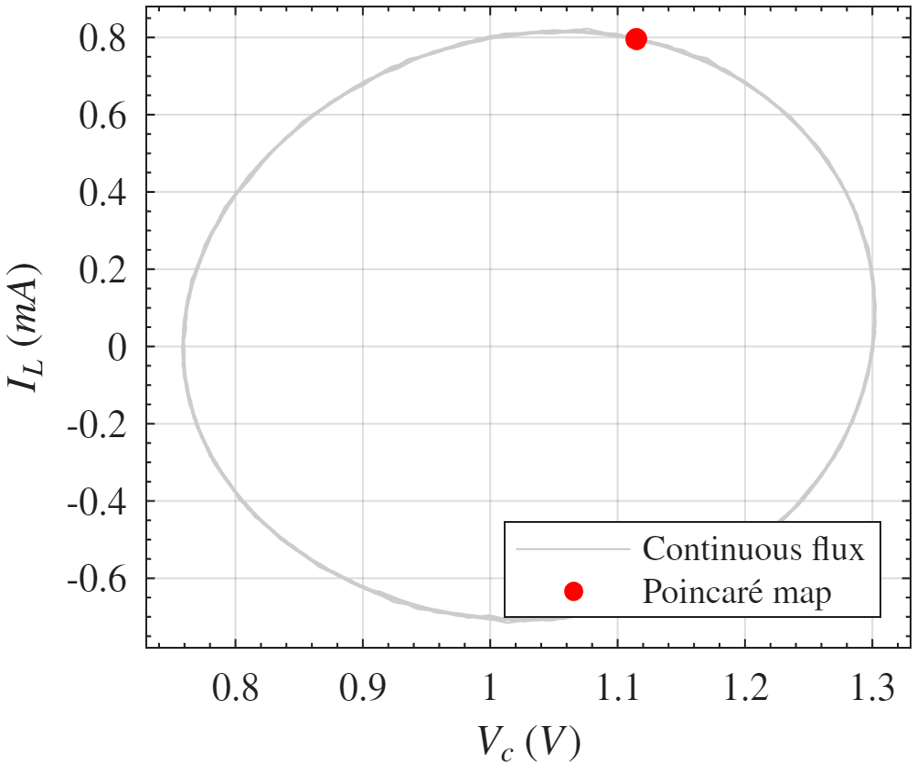}
\end{subfigure}
\begin{subfigure}{1\textwidth}
    \centering
    \caption{}
    \includegraphics[width=0.45\linewidth]{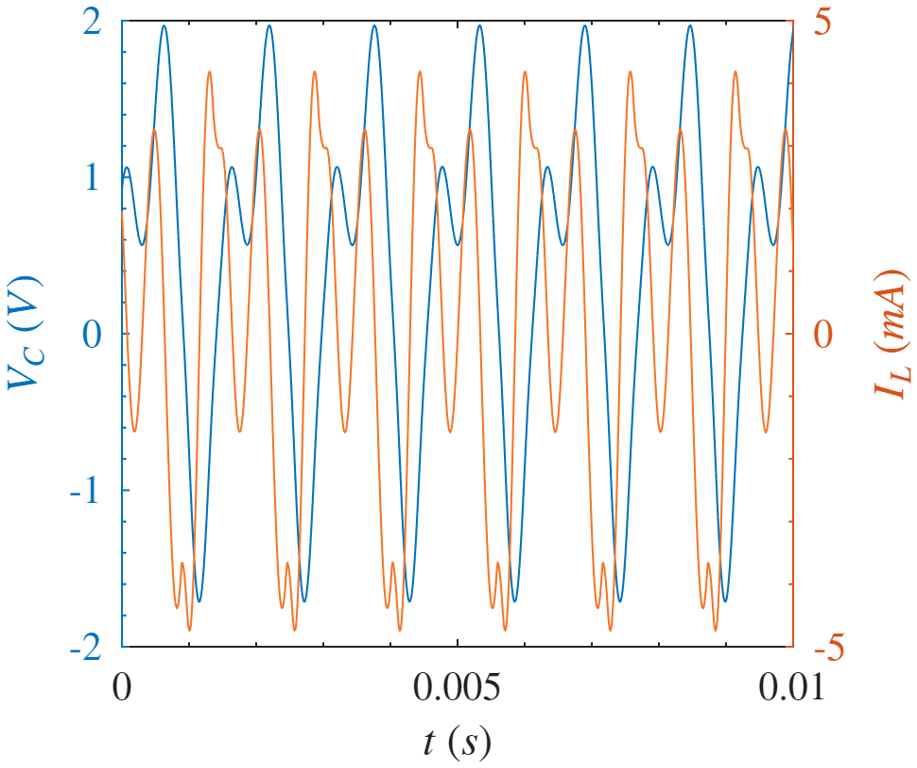}
    \includegraphics[width=0.45\linewidth]{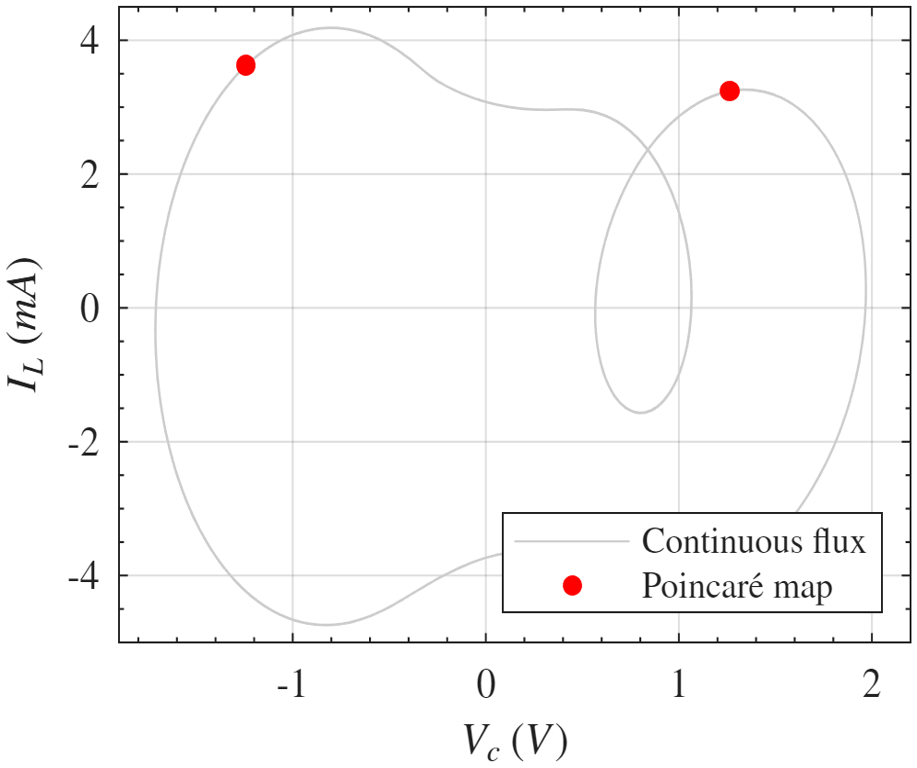}
\end{subfigure}
\begin{subfigure}{1\textwidth}
    \centering
    \caption{}
    \includegraphics[width=0.45\linewidth]{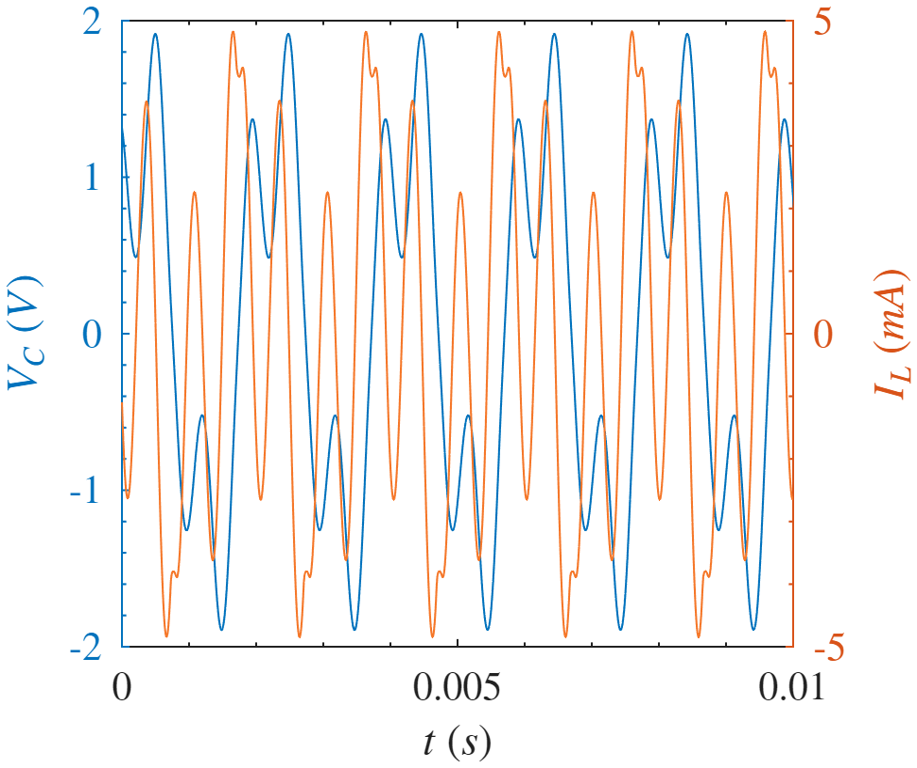}
    \includegraphics[width=0.45\linewidth]{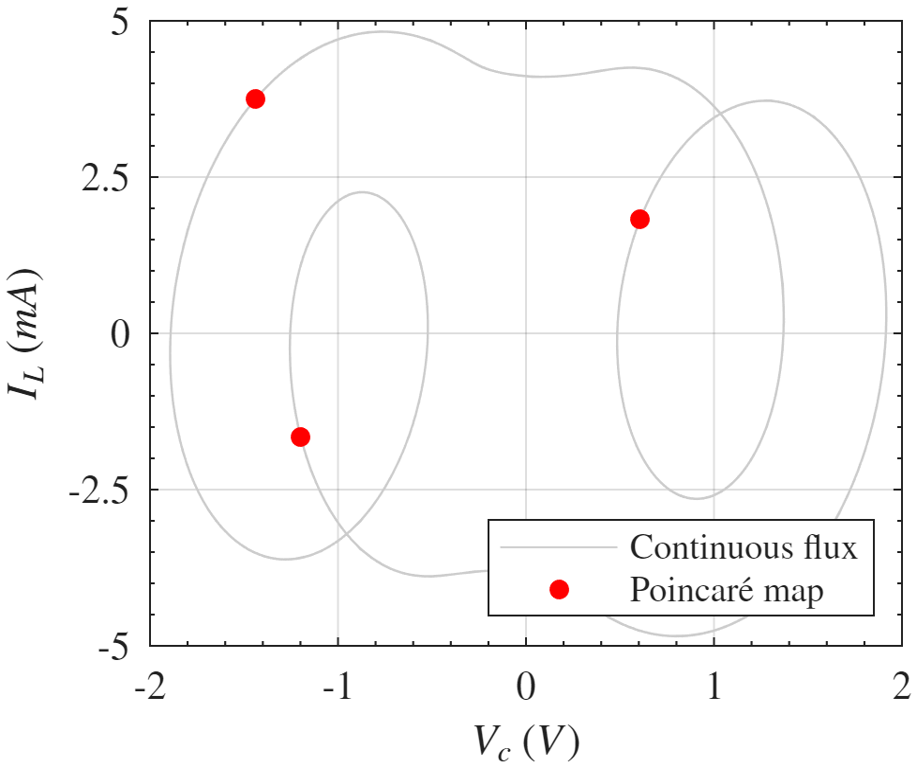}
\end{subfigure}
\caption{Three  examples of periodic behavior.
Capacitor voltage and inductor current time series (left) and phase-space diagram with Poincaré map (right) for different periodic solutions. (a) Period-1 dynamics for $A = 0.155$ V and $f = 940$ Hz, confirmed by a single point in the Poincaré map; the orbit remains confined to one potential well. (b) Period-2 dynamics for $A = 0.512$ V and $f = 1276$ Hz, identified by two points in the Poincaré map. (c) Period-3 dynamics for $A = 0.554$ V and $f = 1514$ Hz, confirmed by three points in the Poincaré map.}
\label{fig:STyMP1}
\end{figure}

\begin{figure}[hbtp]
\begin{subfigure}{1\textwidth}
  \centering
  \includegraphics[width=0.45\linewidth]{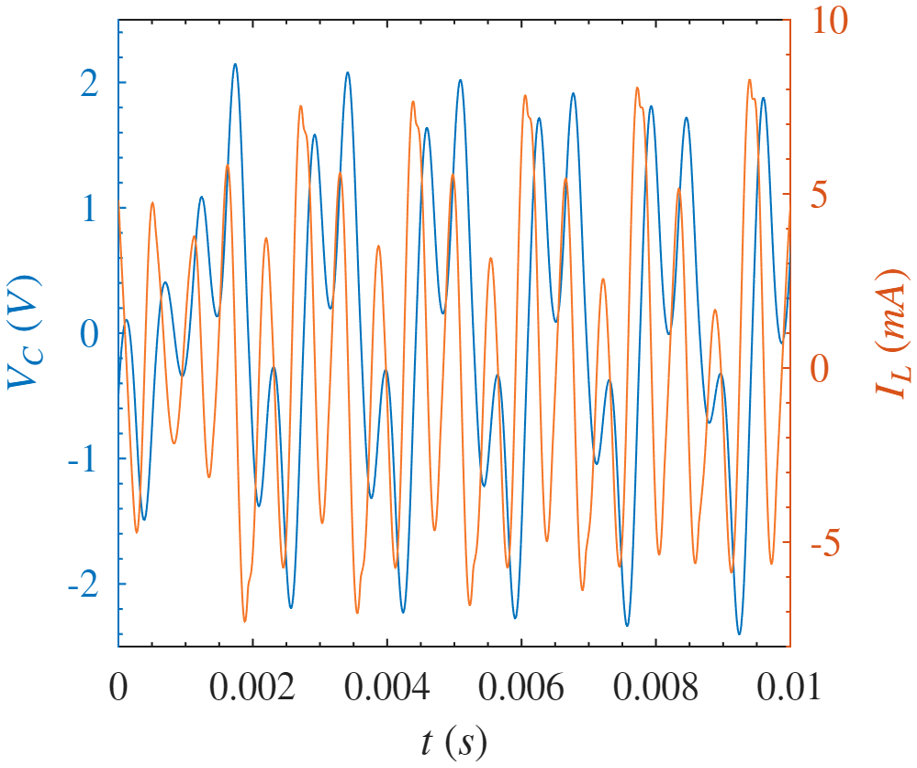}
  \includegraphics[width=0.45\linewidth]{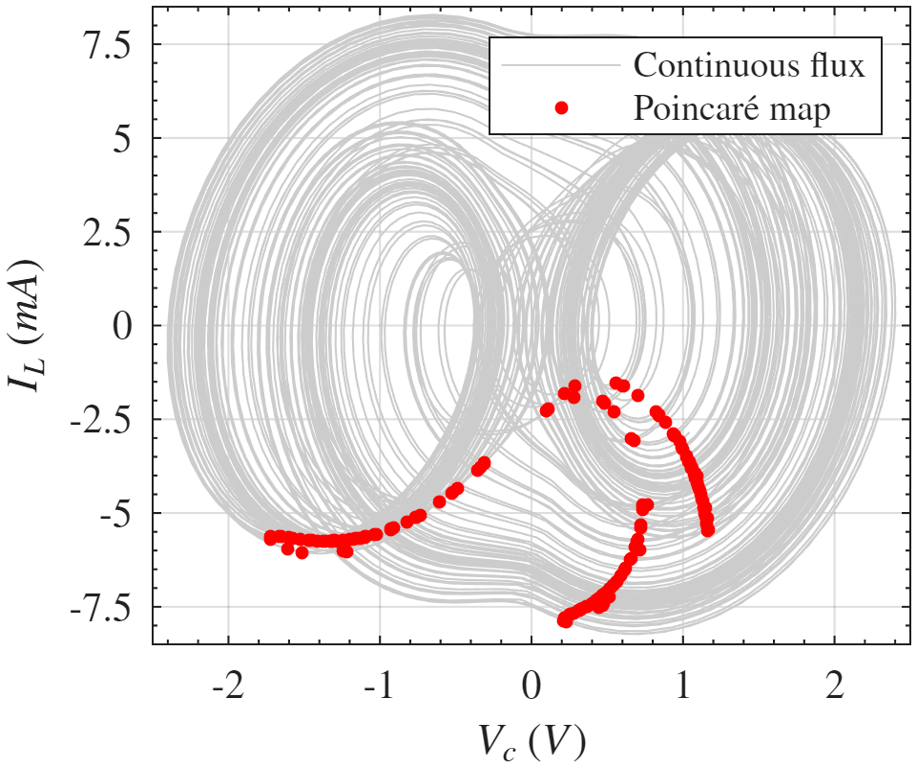}
\end{subfigure}
\caption{Example of chaotic dynamics in the circuit.
Capacitor voltage and inductor current time series (left) and phase-space diagram with Poincaré map (right) with $A = 1.037$~V and
$f = 1794$~Hz. The time series, phase portrait, and Poincar\'e map are all consistent with chaotic dynamics, confirmed by a positive largest Lyapunov exponent.}
\label{fig:STyMP2}
\end{figure}

The experimental results reveal a striking diversity of dynamical 
behaviors as the driving amplitude $A$ and frequency $f$ are varied. 
Figures~\ref{fig:STyMP1} and~\ref{fig:STyMP2} show representative 
time series, phase-space trajectories, and Poincar\'e maps spanning 
the main dynamical regimes of the system. Figure~\ref{fig:STyMP1} 
presents examples of periodic behavior, whereas Fig.~\ref{fig:STyMP2} 
illustrates chaotic dynamics.

Poincar\'e maps were constructed via stroboscopic sampling, recording 
the state variables $(V_C, I_L)$ at discrete times synchronized with 
the external forcing period $T = 1/f = 2\pi/\omega$. This reduction 
of the continuous-time flow to a discrete map sharpens the 
identification of bifurcations and the onset of chaos.

Together, these cases span the principal dynamical regimes of the 
system: periodic confinement within a single potential well (period-1), 
period-doubled motion (period-2), a period-3 window embedded within a 
chaotic region, and fully developed chaos. In each regime the 
Poincar\'e map provides an unambiguous topological signature---a single 
fixed point, two points, three points, or a diffuse strange-attractor 
cross-section---that identifies the dynamical state without relying on 
long-time integration or statistical criteria. 
\subsection{Bifurcation diagrams and Lyapunov exponents}

\begin{figure}[H]
  \centering
  \includegraphics[width=1\linewidth]{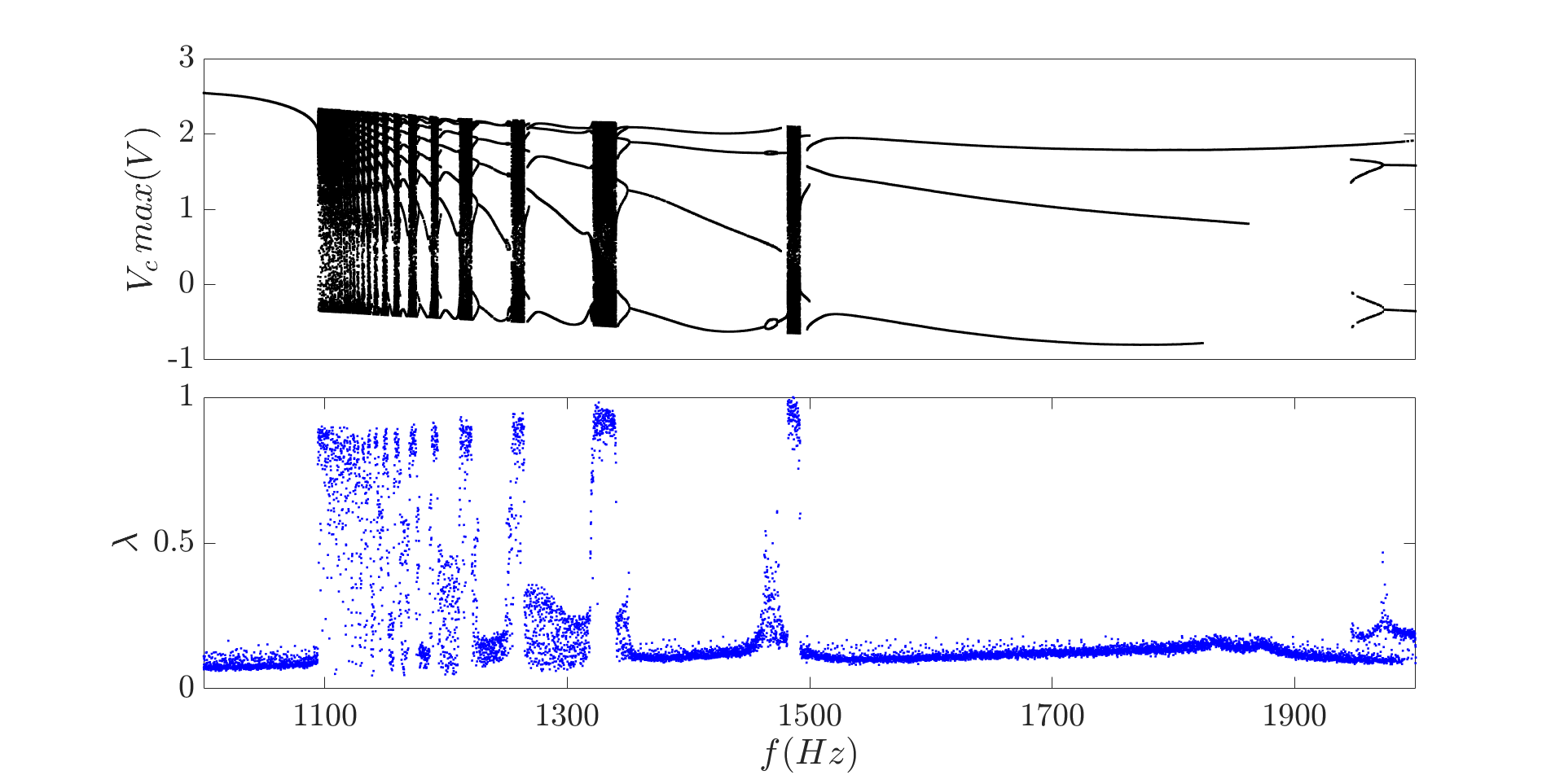}
\caption{Bifurcation diagram (top) for a frequency sweep at fixed amplitude
$A = 0.6065$~V, and the corresponding normalized largest Lyapunov exponent $\lambda$
as a function of frequency (bottom). Positive values of $\lambda$ confirm
the chaotic nature of the aperiodic windows.}
\label{fig:DBA}
\end{figure}

\begin{figure}[H]
  \centering
  \includegraphics[width=1\linewidth]{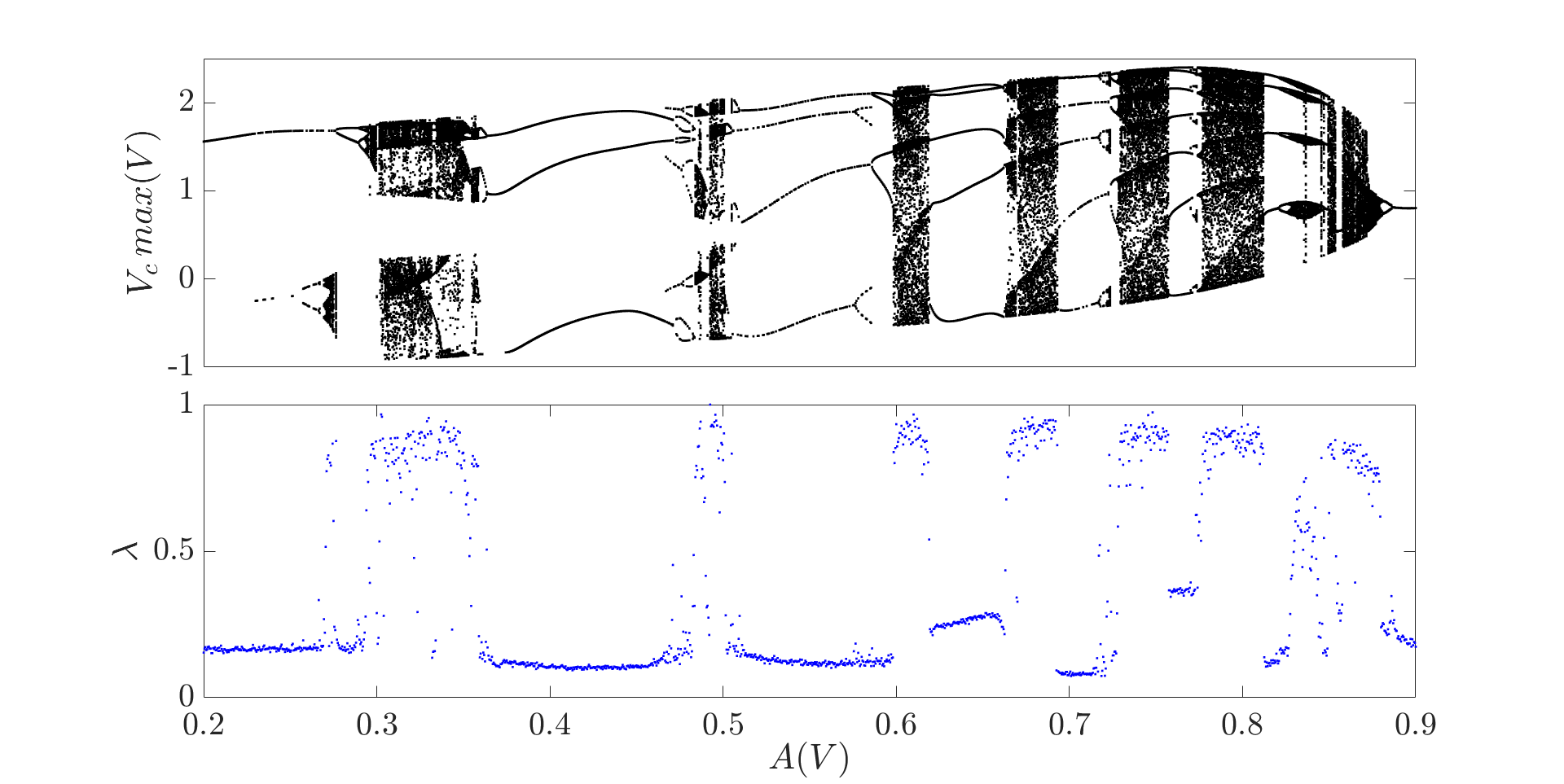}
\caption{Bifurcation diagram (top) for an amplitude sweep at fixed frequency
$f = 1304$~Hz, and the corresponding normalized largest Lyapunov exponent $\lambda$ as a function of amplitude (bottom).}
\label{fig:DBF}
\end{figure}

To characterize the global dynamics, parametric sweeps were performed and
bifurcation diagrams constructed from the local maxima of the recorded time
series. The largest Lyapunov exponent $\lambda$ was computed simultaneously,
providing a quantitative complement to the topological information of the
bifurcation diagrams. Results are presented in Figs.~\ref{fig:DBA}
and~\ref{fig:DBF}.

Figure~\ref{fig:DBA} shows the frequency sweep at fixed amplitude
$A = 0.6065$~V. The system begins in a period-1 regime, maintained up to
$f \approx 1092$~Hz, beyond which successive period-doubling cascades lead
to chaotic windows. The system then restabilizes into a period-3 window
above $f \approx 1500$~Hz. The Lyapunov exponent (lower panel) tracks these
transitions sharply, taking unambiguously positive values in the chaotic
regions and returning to zero at the boundaries of periodic windows.

Also in Fig.~\ref{fig:DBA}, near $f = 1946$~Hz, the bifurcation 
diagram exhibits discontinuous branches---referred to here as 
multistability-induced discontinuities---which are not noise artifacts 
but a direct signature of coexisting attractors~\cite{gutierrez2020observation}. 
Because each frequency value was measured from an independent 
realization rather than by quasi-static continuation, the electrical 
initial conditions vary randomly between realizations. In a bistable 
system, these random initial conditions cause the trajectory to settle 
into different coexisting attractors unpredictably, producing a 
superposition of solution branches in the bifurcation diagram. The 
same mechanism accounts for the apparent period-4 window between 
$f = 1342$~Hz and $f = 1480$~Hz: the circuit is in fact oscillating 
in a period-2 regime, but the stochastic alternation between two 
coexisting period-2 attractors produces a pseudo-period-4 signature 
when their maxima are superimposed.

Figure~\ref{fig:DBF} shows the amplitude sweep at fixed frequency
$f = 1304$~Hz, revealing multiple periodic windows, period-doubling cascades,
and broad chaotic regions, whose aperiodic nature is confirmed by large
positive values of $\lambda$ in the lower panel.

It is worth noting that due to the experimental nature of the time series and finite-time computation constraints, the largest Lyapunov exponent exhibits a positive baseline shift even within periodic windows, as typically expected from noise-contaminated data. Nevertheless, the qualitative trend of the exponent remains highly consistent with the bifurcation diagram, showing sharp drops during periodic behavior and pronounced peaks in chaotic regimes.
\subsection{Parameter-space diagram}
\label{sec:espacio_parametros}

\begin{figure}[H]
  \centering
  \includegraphics[width=1\linewidth]{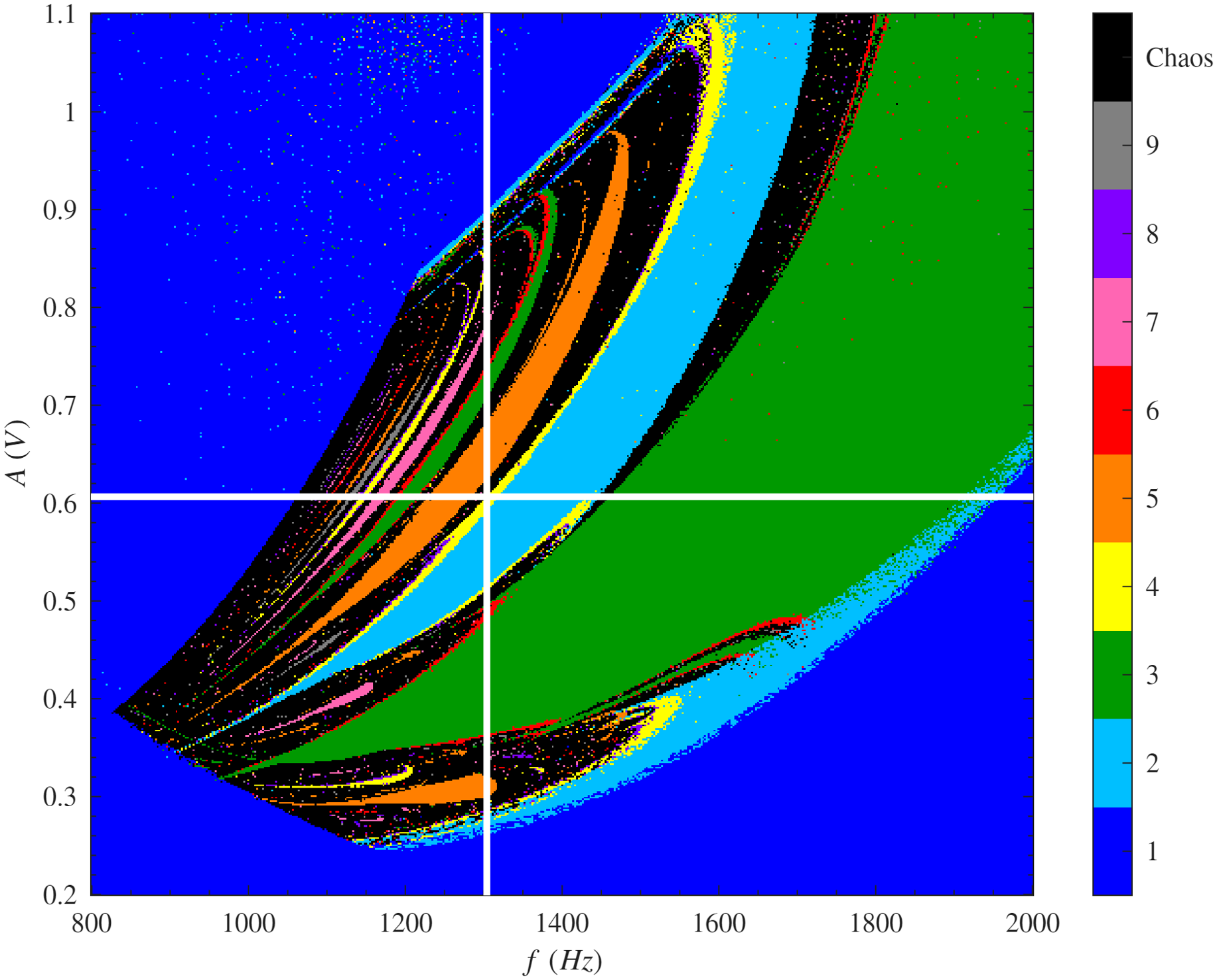}
\caption{Experimentally obtained dynamical phase diagram in the $(f,A)$
parameter space ($500\times500$ points). Colors indicate the periodicity
of the response; black regions correspond to chaotic dynamics ($\lambda > 0$).
The white horizontal and vertical lines indicate the parameter sweeps 
presented in Figs. \ref{fig:DBA} and \ref{fig:DBF}, respectively.}
\label{fig:EspacioParam}
\end{figure}

A global portrait of the dynamics was obtained by constructing a
high-resolution diagram in the two-dimensional parameter space $(f, A)$,
shown in Fig.~\ref{fig:EspacioParam}. Each of the $500 \times 500$ measured
points is color-coded according to the periodicity of the response,
determined from the number of Poincar\'e map intersections; black regions
denote chaotic dynamics, validated by $\lambda > 0$.

The diagram reveals with remarkable clarity the intricate morphology of
the boundaries between periodic and chaotic regions. Broad low-period
domains coexist with narrow periodic windows embedded within chaotic
seas, and complex isoperiodic structures, reminiscent of Arnold tongues
and shrimp-shaped sequences~\cite{gallas1993structure,cabeza2013periodicity}, emerge from the
chaotic background with well-defined boundaries. This analog implementation captures the fractal nature of these boundaries and detects narrow periodic windows within chaotic regions, features that are otherwise difficult to resolve, thereby demonstrating the high precision of the measurement system.

\subsection{Antiperiodicity}
\label{sec:antiperiodicidad}

A notable feature of the observed dynamics is the presence of antiperiodic
orbits, a property intrinsic to systems with symmetric double-well potentials.
An orbit is antiperiodic if it satisfies

\begin{equation}
    x(t + T/2) = -x(t),
    \label{eq.antip}
\end{equation}
where $T$ is the forcing period \cite{freire2013antiperiodic,freire2014self,shaw2015antiperiodic}. Geometrically, an antiperiodic trajectory
is invariant under a $180^\circ$ rotation in phase space and therefore visits
both potential wells symmetrically, whereas a non-antiperiodic orbit breaks
this symmetry and remains predominantly confined to one well.

\begin{figure}[H]
\captionsetup[subfigure]{labelformat=empty}
\centering
\begin{subfigure}{.32\textwidth}
  \centering
  \includegraphics[width=\linewidth]{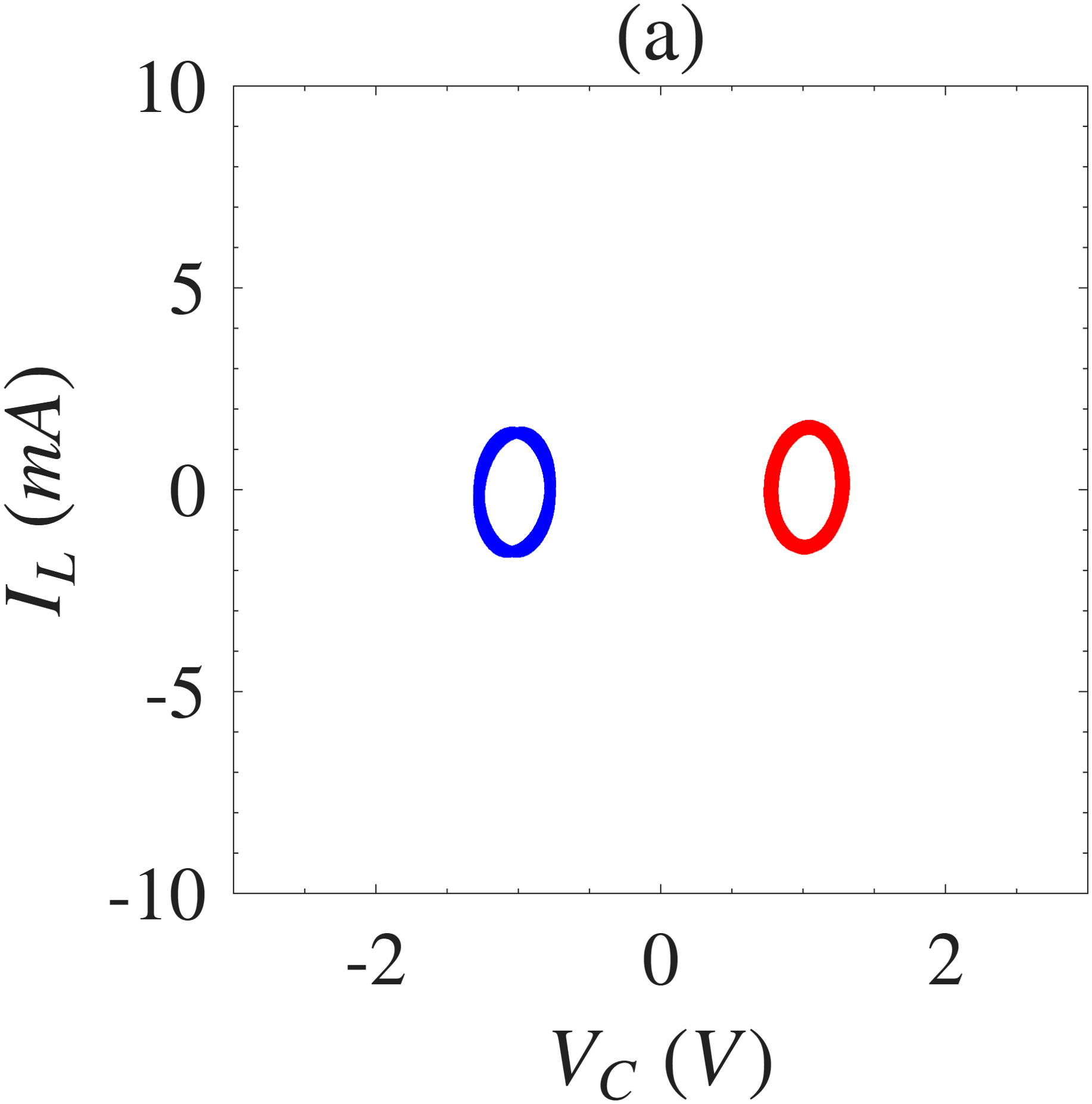}
  \caption{}
  \label{fig.antip1}
\end{subfigure}
\begin{subfigure}{.32\textwidth}
  \centering
  \includegraphics[width=\linewidth]{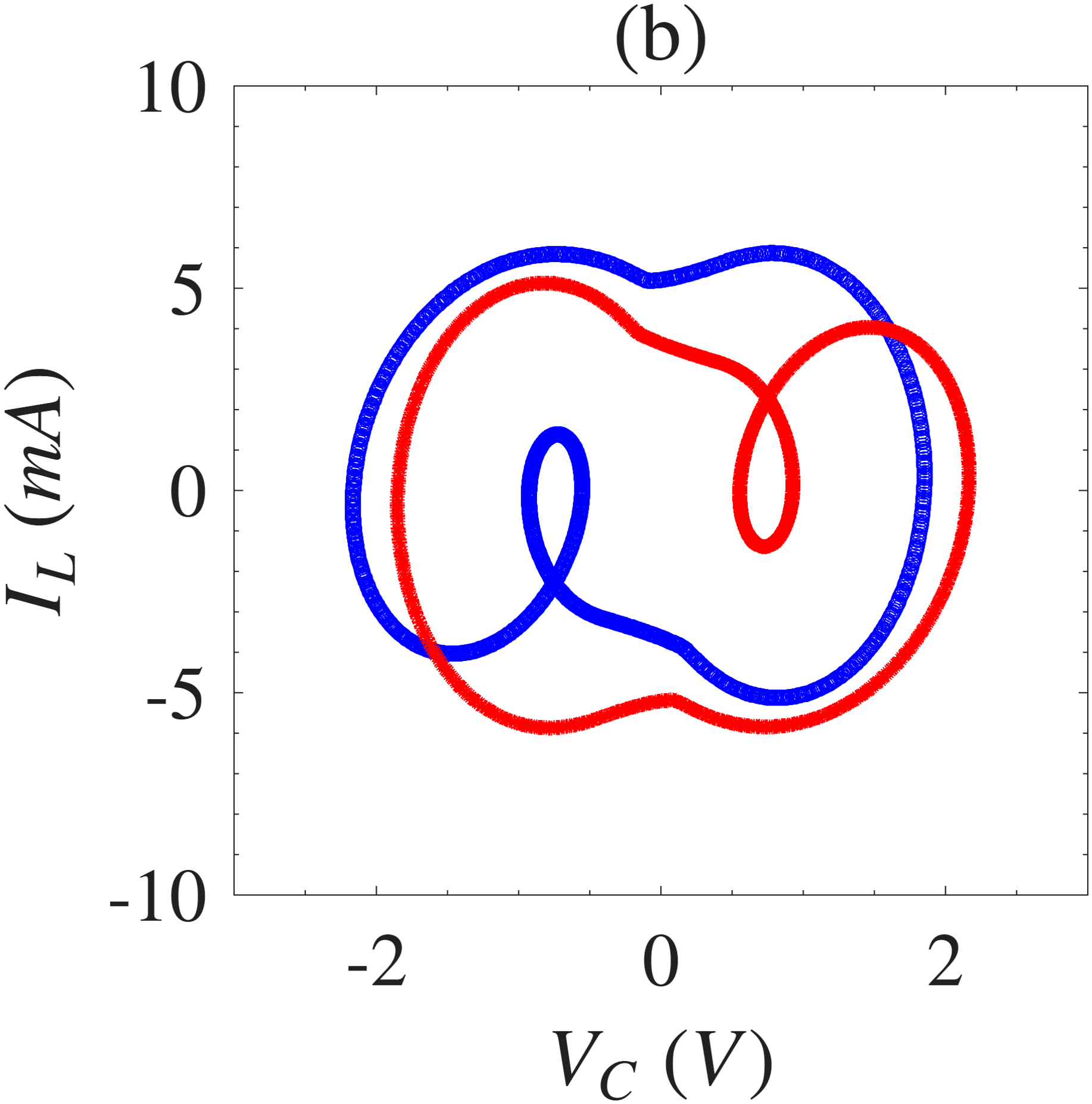}
  \caption{}
  \label{fig.antip2}
\end{subfigure}
\begin{subfigure}{.32\textwidth}
  \centering
  \includegraphics[width=\linewidth]{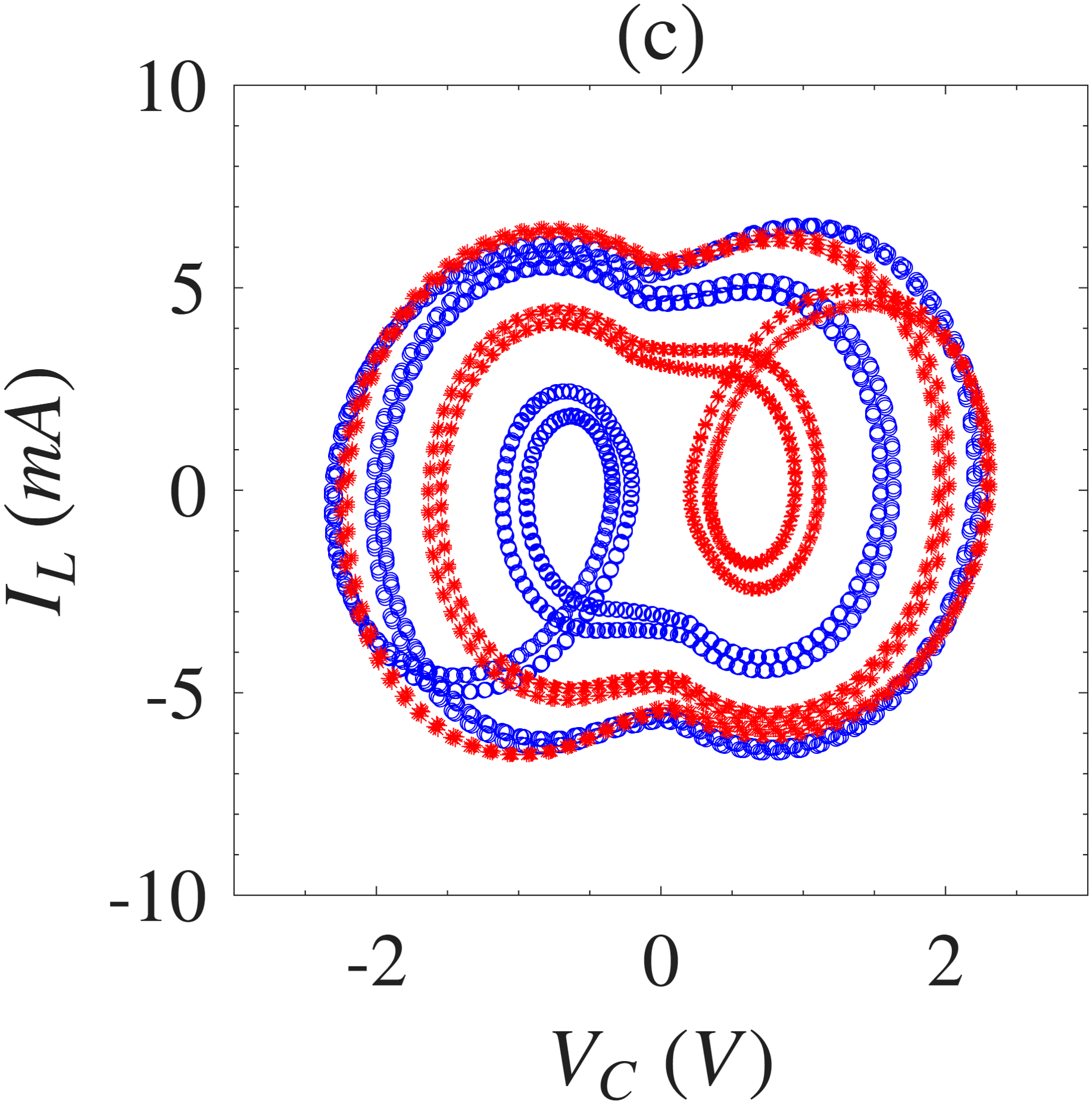}
  \caption{}
  \label{fig.antip6}
\end{subfigure}
\par\medskip
\begin{subfigure}{.32\textwidth}
  \centering
  \includegraphics[width=\linewidth]{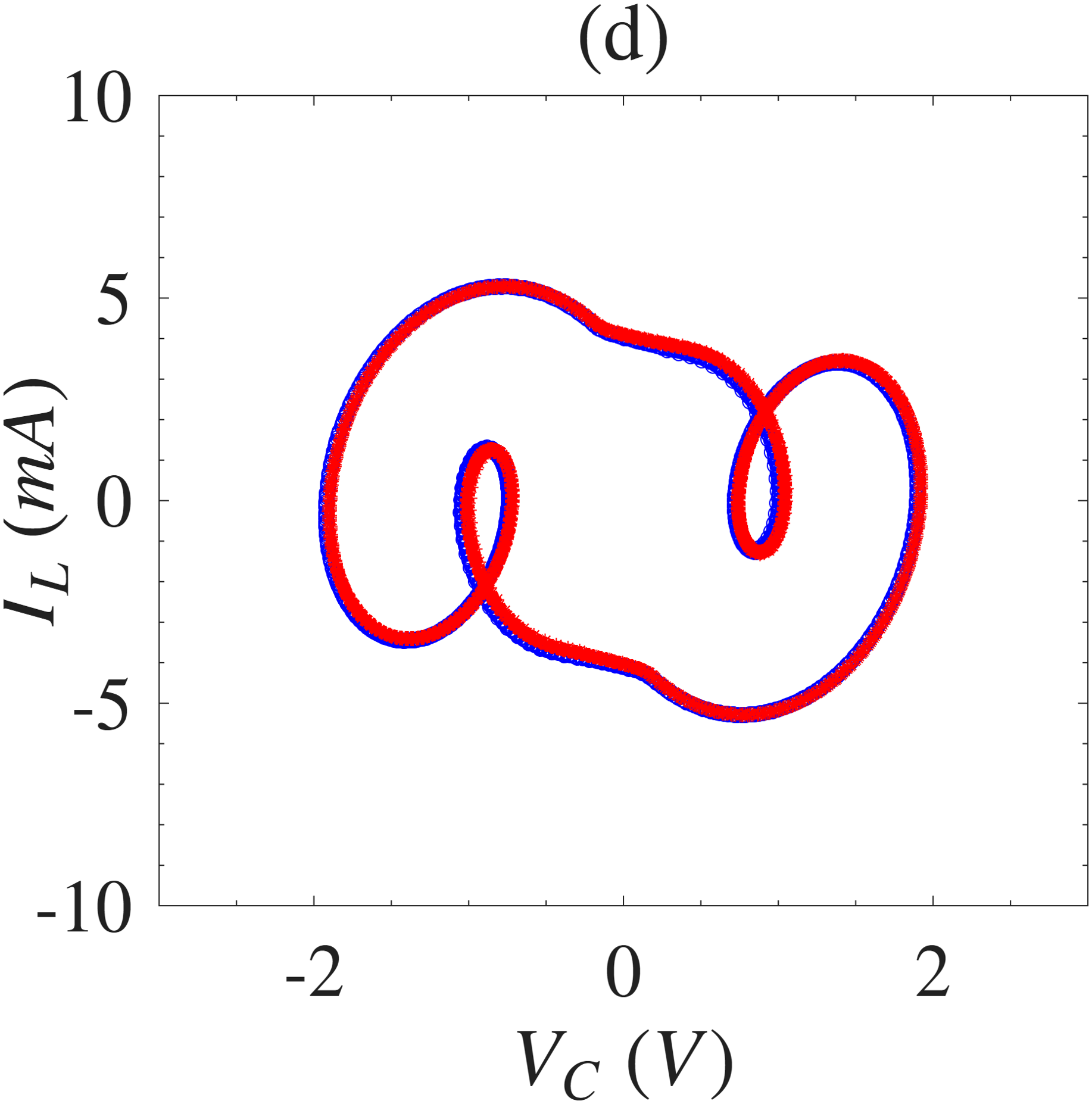}
  \caption{}
  \label{fig.antip3}
\end{subfigure}
\begin{subfigure}{.32\textwidth}
  \centering
  \includegraphics[width=\linewidth]{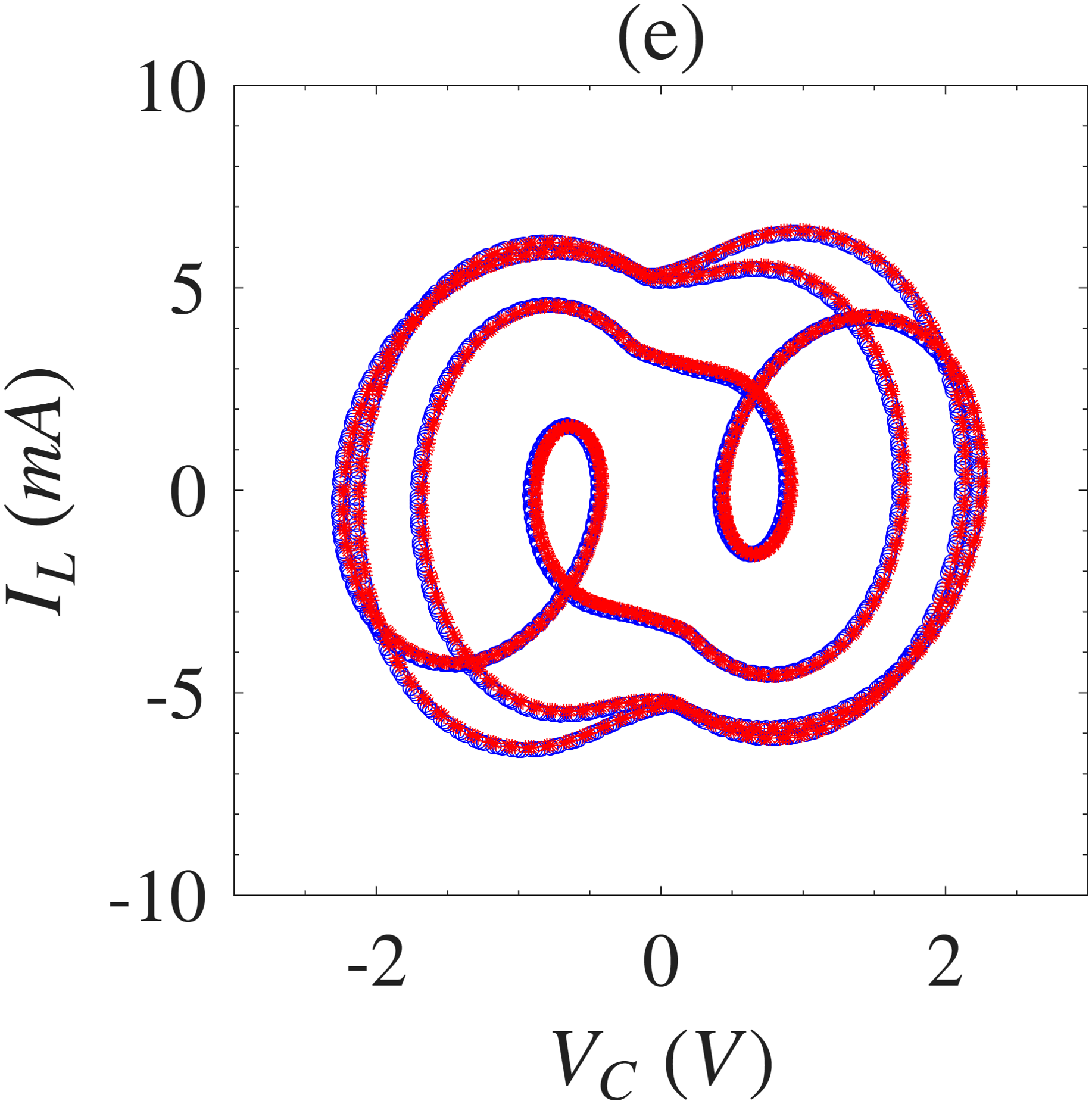}
  \caption{}
  \label{fig.antip5}
\end{subfigure}
\begin{subfigure}{.32\textwidth}
  \centering
  \includegraphics[width=\linewidth]{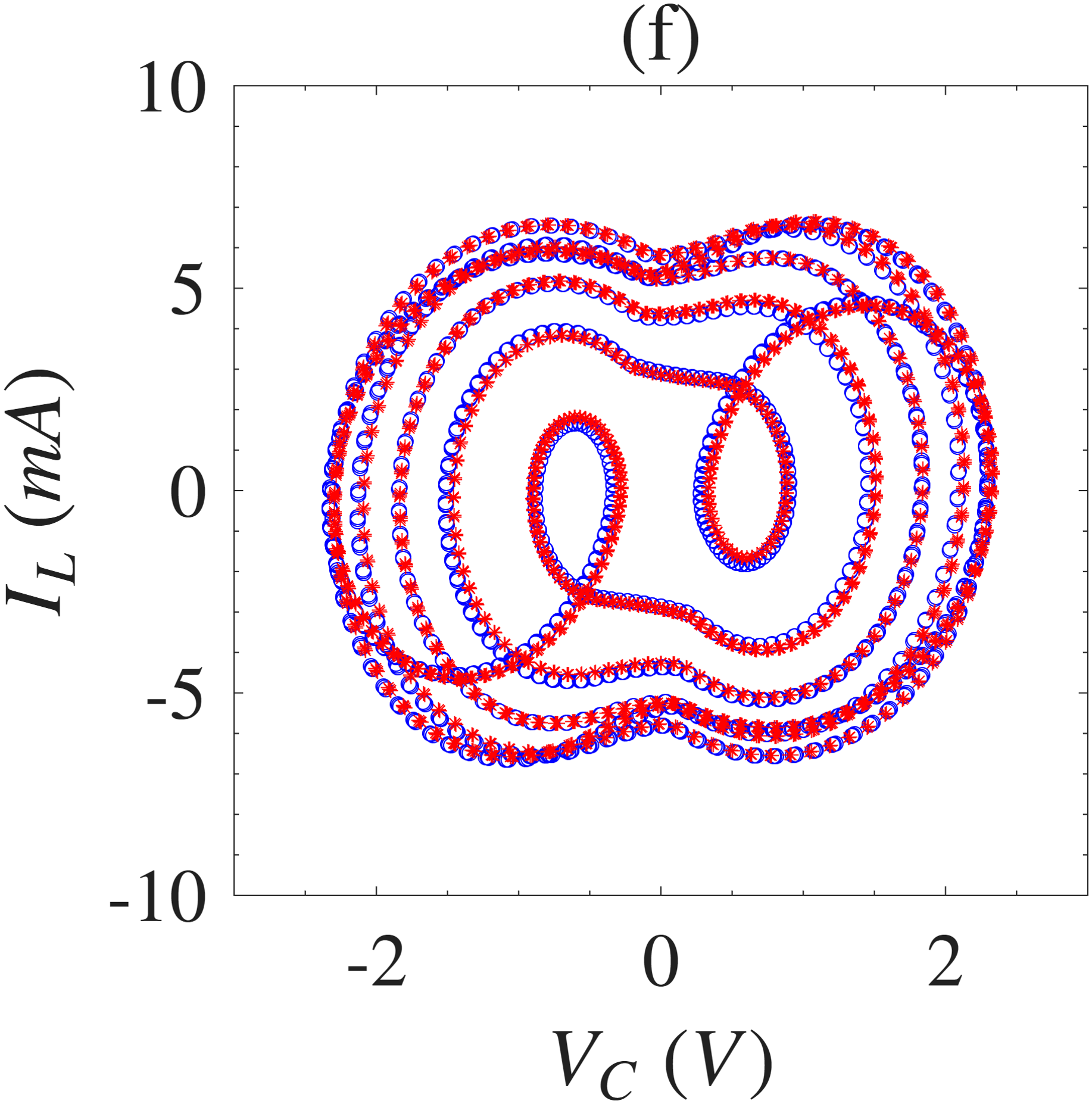}
  \caption{}
  \label{fig.antip7}
\end{subfigure}
\caption{Phase portraits illustrating the transition between symmetry-broken
and antiperiodic regimes. In each panel, the experimental orbit (blue) is
shown together with its $180^\circ$-rotated counterpart (red).
(a)~Period-1: symmetry breaking; the orbit is confined to one well and does
not overlap with its rotation. (b)~Period-2 and (c)~Period-6: asymmetric
attractors of increasing topological complexity, requiring the rotated
counterpart to reconstruct the global symmetry. (d)~Period-3,
(e)~Period-5, and (f)~Period-7: intrinsically antiperiodic odd-period
orbits satisfying Eq.~\eqref{eq.antip}; the blue and red curves overlap
completely.}
\label{fig.antip}
\end{figure}

Figure~\ref{fig.antip} illustrates the interplay between symmetry breaking
and symmetry restoration across different dynamical regimes. In the
period-1 case (panel~a), the orbit has lost antiperiodicity and remains
confined to one potential well; applying a $180^\circ$ rotation reveals the
coexisting twin attractor (red curve), a hallmark of the symmetry-broken
regime that precedes the period-doubling cascade. As the system follows this
cascade, the period-2 and period-6 attractors (panels~b and~c) maintain
their asymmetry while acquiring additional topological complexity.

In contrast, the odd-period windows (panels~d--f) correspond to intrinsically
antiperiodic orbits that satisfy Eq.~\eqref{eq.antip} exactly: the blue
trajectory is invariant under rotation, producing a full overlap with the
red curve. These orbits cross the potential barrier and visit both wells
symmetrically, representing islands of global-symmetry restoration embedded
within the chaotic sea. The ability to resolve the period-7 orbit
(panel~f) with the observed sharpness confirms that the experimental
resolution is well above the thermal noise floor of the components, validating
the fidelity of the instrumentation described in Section~\ref{sec.montaje}.

\section{Conclusions and perspectives}
\label{sec:conclusions}

An analog electronic circuit for the experimental study of the Duffing--Holmes oscillator has been implemented and characterized. By means of systematic sweeps of the driving amplitude and frequency, a comprehensive real-time, high-resolution exploration of the system's nonlinear dynamics has been carried out, providing detailed insight into the diverse dynamical regimes and phenomena  exhibited by the oscillator.

The reconstruction of phase portraits and Poincar\'e maps from experimental 
time series allowed rigorous identification and classification of the 
observed dynamics. Beyond the canonical period-doubling route to chaos 
--- corroborated quantitatively by the largest Lyapunov exponent spectrum 
--- the experiments revealed several features of particular interest. 
Antiperiodic orbits of odd period ($3T$, $5T$, and $7T$) were identified, 
in which the trajectory satisfies $x(t + T/2) = -x(t)$ and thus recovers 
the full symmetry of the double-well potential. The methodology of 
independent realizations exposed multistability through spontaneous jumps 
between coexisting attractors and the emergence of ghost bifurcations, 
while the documentation of asymmetric twin attractors illustrated how 
local symmetry breaking can coexist with the global symmetry of the 
system. Throughout, the close agreement between continuous phase portraits, 
the discrete topology of the Poincar\'e maps, and the Lyapunov exponent 
calculations validated both the precision of the acquisition system and 
the fidelity of the circuit to the underlying mathematical model.

A central experimental contribution of this work is the construction of 
a high-resolution dynamical diagram in the two-dimensional parameter 
space $(f,A)$. This map delineates with remarkable clarity the boundaries 
between periodic and chaotic regimes and reveals the emergence of complex 
isoperiodic structures, reminiscent of Arnold tongues and shrimp-shaped 
sequences~\cite{gallas1993structure}, embedded within broad chaotic 
regions. Beyond its descriptive value, this diagram demonstrates that 
analog hardware 
provides a high-throughput, continuous-time platform capable of efficiently mapping vast multidimensional parameter spaces and resolving fine-scale topological structures with high experimental fidelity.

Several directions remain open for future investigation. On the 
theoretical side, a rigorous mapping of the basins of attraction and a 
systematic characterization of the boundary crises associated with 
intermittency would complement the bifurcation analysis presented here. 
The antiperiodic orbits documented in this work, particularly the 
high-order $7T$ cycle, raise a natural question about the robustness 
of the symmetry condition $x(t + T/2) = -x(t)$ against the residual 
hardware asymmetries inevitable in physical components, and whether its 
breakdown can serve as an early precursor of attractor crises or 
intermittent transitions. On the applied side, the flexibility and 
robustness of the circuit design open direct avenues toward the 
implementation of chaotic synchronization schemes with potential 
applications in secure communications.

\section*{Declaration of Generative AI and AI-Assisted Technologies in the Manuscript Preparation Process}

During the preparation of this work the authors used Claude 
(Anthropic, claude.ai) in order to improve the clarity, style, and 
academic register of the written English. After using this tool, the 
authors reviewed and edited the content as needed and take full 
responsibility for the content of the published article.

\bibliographystyle{elsarticle-num}
\bibliography{sample} 

\newpage
\section*{Appendix: Electronic Implementation Details}
\label{sec:Anexo}

\begin{figure}[H]
  \centering
  \includegraphics[width=.5\linewidth]{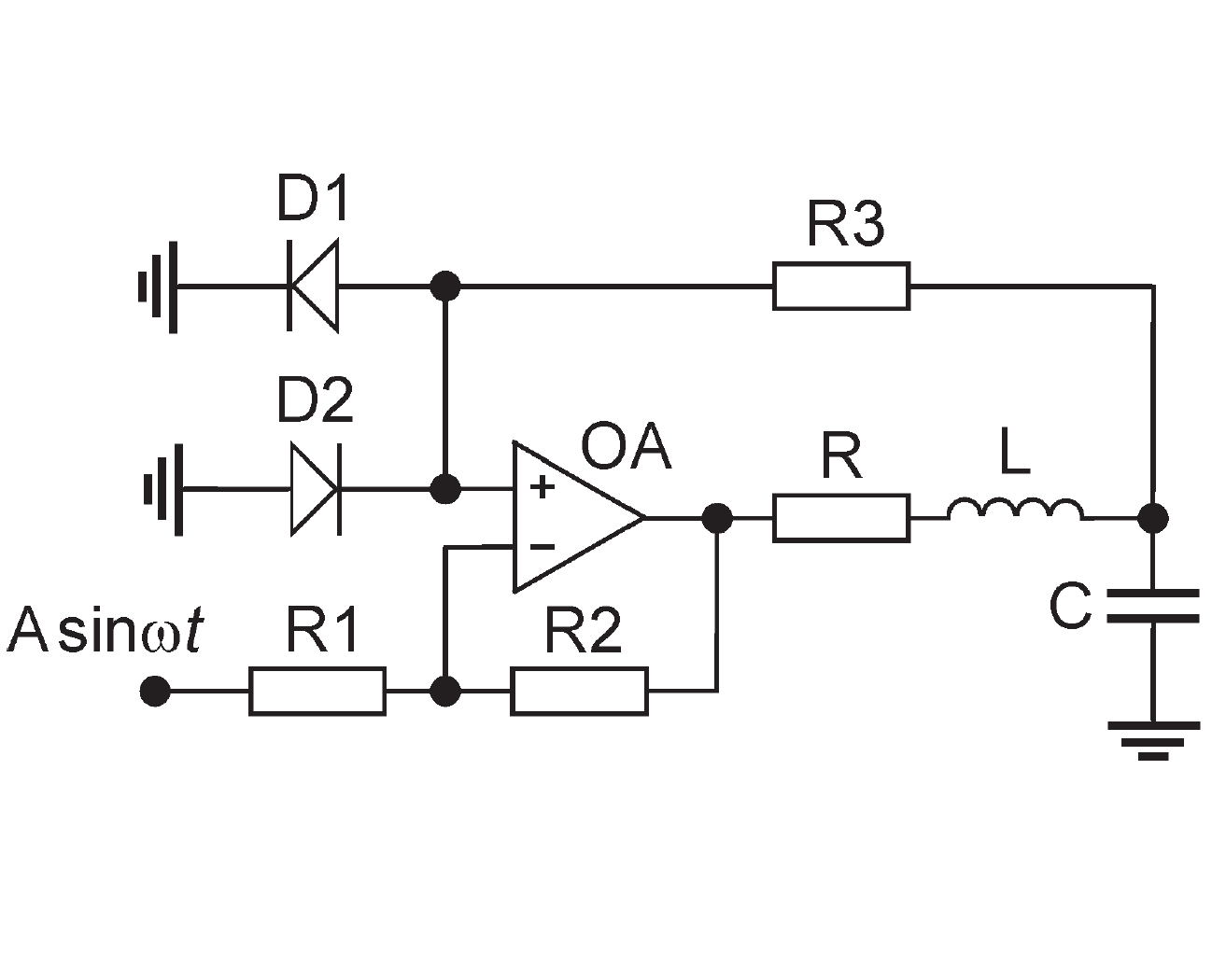}    \   \   \  \   \includegraphics[width=0.35\linewidth]{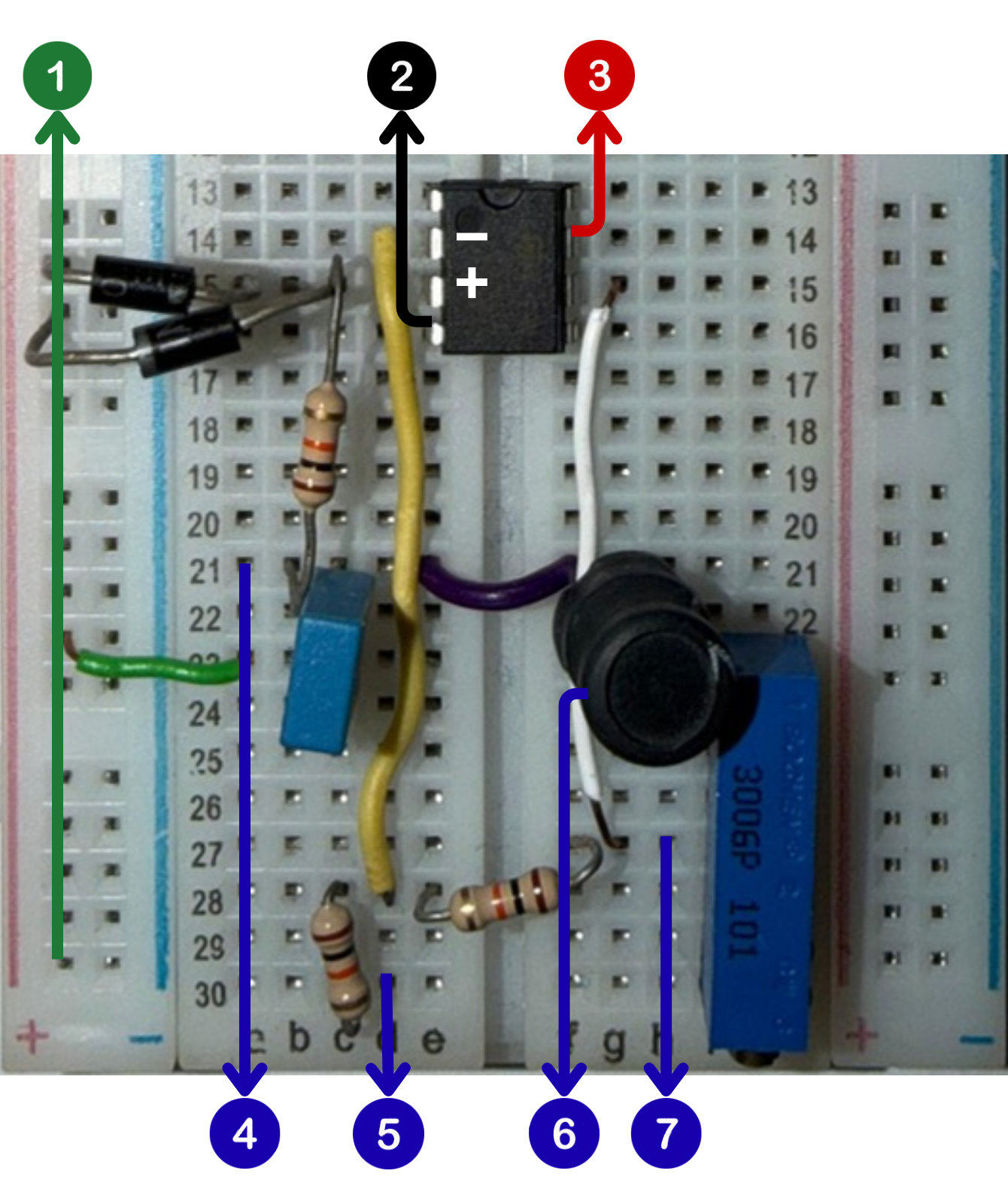}
\caption{(Left) Schematic of the implemented circuit ~\cite{tamaseviciute2008analogue}.  (Right) Photograph that illustrates the physical placement of the components on the breadboard and their respective connections: \textcircled{1} Grounded line, \textcircled{2} Operational amplifier negative supply voltage, \textcircled{3} Operational amplifier positive supply voltage, \textcircled{4} Capacitor's voltage ($V_C$) measured through the data acquisition card, \textcircled{5} Forcing signal input, and \textcircled{6}, \textcircled{7} Resistor differential voltage measured through the data acquisition card to determine the current ($I_L$).}
\label{fig:Montaje}
\end{figure}

To ensure the experimental reproducibility of the described chaotic and multistable phenomena, this appendix provides a detailed description of the physical hardware assembly. Figure \ref{fig:Montaje} shows the complete circuit schematic along with a high-resolution photograph illustrating the spatial arrangement of the components and the mounting strategy on the breadboard. This specific arrangement was carefully optimized to minimize parasitic capacitances and electromagnetic pickup, thereby preserving the integrity of the analog computing core. The nominal values and experimental uncertainties of the passive components and diode threshold voltages are summarized in Table \ref{tab:componentes}. 

\begin{table}[h]
\centering
\caption{Specifications, type and experimental uncertainties of the electronic components.}
\label{tab:componentes}
\definecolor{grisclaro}{rgb}{0.95, 0.95, 0.95}
\rowcolors{2}{white}{grisclaro}
\label{tab:components}
\begin{tabular}{llll}
\hline
\textbf{Component} & \textbf{Type} & \textbf{Nominal Value} & \textbf{Unit} \\ \hline
$R_1 = R_2 = R_3$   & Metal Film            & $10.0 \pm 0.1$         & $\text{k}\Omega$ \\
$R$                 & Potentiometer (tuned)& $20.0 \pm 0.4$         & $\Omega$ \\
$C$                 & Polyester Film        & $0.47 \pm 0.02$        & $\mu\text{F}$ \\
$L$                 & Radial Inductor    & $22.3 \pm 0.7$         & $\text{mH}$ \\
$D_1, D_2 \ \   (V_{\text{th}})$ &Diodes  1N4007  &  $0.58 \pm 0.01$ & $\text{V}$ \\
Op-Amp ($OA$)       & UA741       & --- & --- 
\\ \hline
\end{tabular}
\end{table}

\end{document}